\documentclass[sigplan,10pt,nonacm]{acmart}
\begin{document}

\title{Nofl: A Precise Immix}

\author{Andy Wingo}
\email{wingo@igalia.com}

\affiliation{%
  \institution{Igalia, S.L.}
  \city{A Coruña}
  \country{Spain}
}

\begin{abstract}
Can a memory manager be built with fast bump-pointer allocation,
single-pass heap tracing, and a low upper bound on memory overhead?  The
Immix collector answered in the affirmative for the first two, but the
granularity at which it reclaims memory means that in the worst case a
tiny object can keep two 128-byte lines of memory from being re-used for
allocation.

This paper takes Immix to an extreme of precision, allowing all free
space between objects to be reclaimed, down to the limit of the
allocator's minimum alignment.  We present the design of this Nofl
layout, build a collector library around it, and build a new Scheme-to-C
compiler as a workbench.  We make a first evaluation of the Nofl-based
mostly-marking collector when compared to standard copying and
mark-sweep collectors and run against a limited set of microbenchmarks,
finding that Nofl outperforms the others for tight-to-adequate heap
sizes.
\end{abstract}

\ccsdesc[500]{Software and its engineering~Garbage collection}
\ccsdesc[300]{Software and its engineering~Runtime environments}

\keywords{memory management, garbage collection, immix, scheme}

\maketitle

\section{Introduction}
Garbage collection is a design space consisting of trade-offs rather
than optima. There are a small number of fundamental algorithms, each
with their own characteristics: \emph{mark-sweep} is parsimonious with
memory, but at the expense of allocation time, fragmentation, and cache
locality; \emph{mark-compact} allows for contiguous allocation and good
cache locality, but at the expense of multiple passes over the heap; and
\emph{copying} has good allocation performance and single-traversal
collection, but at the expense of needing more memory. A given collector
implementation will be built by composing these algorithms together in
different doses.

The Immix algorithm \cite{Immix08} is a design point between mark-sweep
and copying.  Immix collection will typically mark objects in place. If
fragmentation exceeds a threshold, Immix will identify a subset of the
heap for evacuation before collection starts.  When the trace visits an
object in that heap subset, Immix will attempt to move that object into
one of the empty blocks that it keeps as an evacuation reserve.  Immix
sizes the subset of the heap subject to evacuation to fill the space
available, based on liveness calculations from the previous collection,
and relying on the ability to fall back to in-place marking if the
evacuation reserve runs out.  In this way Immix can defragment with only
a few percent of heap space kept in reserve for evacuation.

A number of garbage collection system designers have incorporated Immix
over the last 15 years \cite{Inko, ScalaNative, patton2023parallel}, as
Immix allows for fast bump-pointer allocation, effectively eliminates
fragmentation, and has low average memory overhead. Additionally,
because Immix does not rely on evacuation of any individual object for
collection to succeed, it is well-adapted to run-times that require the
ability to pin some objects in place, either for the rest of the
object's lifetime (e.g.~because object's address was exposed in some
way) or temporarily (e.g.~because the object was passed to a foreign
function). Finally, Immix has served as an adaptable substrate for
experimentation, for example to add conservative root-finding
\cite{shahriyar2014fast}, in-place generational collection
\cite{shahriyar2014fast}, and even reference counting \cite{RCImmix13,
  LXR22}.

This paper takes Immix in a new direction. Our contributions are (1) the
design of a \emph{Nofl space}, consisting of a novel combination of
Immix-inspired optimistic evacuation with a side table of mark bytes,
and (2) an implementation of an embeddable memory management library
containing a Nofl-based mostly-marking collector, as well as a
conventional parallel copying collector and a shim to the
Boehm-Demers-Weiser conservative collector, for comparison. Using the
other collectors as performance oracles, we make an initial evaluation
of the mostly-marking collector with respect to a handful of
microbenchmarks; a proper evaluation will require integrating the
mostly-marking collector into a production language implementation,
which is an ongoing effort.

\section{Deriving Nofl}

Let us begin by examining the mechanism that Immix uses to identify
spans of unused memory to use for bump-pointer allocation: the
\emph{line mark table}. An Immix heap is structured into 32-kilobyte
blocks, each of which is broken into 128-byte \emph{lines}. Objects may
span multiple lines; objects larger than 8 kB are allocated in a
separate space. Each line has a mark byte, in addition to the mark bit
associated with each object.

During collection, as each object is marked, Immix arranges to also mark
the lines that it is on. Under Immix, objects are required to have a bit
indicating whether they are \emph{small} or \emph{medium}; small objects
are smaller than a line, and so Immix just marks the line containing the
object's address and, as a conservative approximation for small objects
spanning a line boundary, the next line as well. Medium objects are
traced to determine their size, and thus how many lines to mark.

At the end of the trace, Immix performs a coarse sweep over each block's
line mark array. Spans of unmarked lines (\emph{holes}) are available
for bump-pointer allocation during the next cycle. Blocks that have no
marked lines are gathered to a separate list, and may be re-used for
allocations, kept as part of the evacuation reserve, or paged out as
large objects are allocated.

\paragraph{Immix line marks are fine-grained but not precise}
The original Immix paper describes the line marking strategy as its
fine-grained unit of reclamation, relative to the coarse blocks which
are the unit of interaction with the operating system.  However, choice
of line size has an uneasy tension with precision.

Immix could have been designed to sweep the allocated memory directly
instead of maintaining a line mark array: the sweeper would traverse
each block, collecting contiguous spans of unmarked objects as
allocatable holes. Relative to direct sweeping, using a line mark array
is less precise---fewer holes will be identified, and those holes will
be smaller---but Immix trades off precision for speed. Indeed in the
original Immix paper, the speed gain is such that sweeping can occur
during the pause; if each object needed to be visited to determine its
length, lazy or concurrent sweeping would have been required.

\paragraph{A side table of object marks can be used as a precise line mark array}
Consider that in some systems, there is already metadata associated with
each location at which an object can start, for example to indicate
start-of-object or to encode a map of pointer fields.  We can use this
metadata to identify holes instead of a more coarse line mark table.

\subsection{Nofl: A Precise Immix}
Nofl is an Immix that reclaims memory precisely by sweeping over a side
table of mark bytes instead of an auxiliary line mark table.  An
equivalent perspective is that Nofl is an Immix whose lines are as small
as possible.  This is a small change with many implications.

\paragraph{Overhead}
Nofl uses one metadata byte per 16 payload bytes, leading to a memory
overhead of 6.25\%. Compare this to Immix's more lightweight overhead of
1/128, or 0.8\%. On the other hand, by representing mark state in a side
table, the embedder no longer needs to reserve space for collector state
in its object representations. Indeed, our Nofl-based collector allows
for pairs in a Scheme system to be represented with just two words,
which is not possible for collectors that require more than two or three
bits of state in object headers. We may thus take 6.25\% to be an
upper-bound estimate of Nofl's total memory overhead.

\paragraph{Extent}
Whereas Immix coarsely marks the extent of a live object via line marks,
Nofl relies on the allocator setting bits in the first and last metadata
byte corresponding to the object's extent.  This allows Nofl to
precisely sweep with a linear scan over bytes, instead of tracing at
sweep-time.  Precision comes at the cost of having to scan more memory
than Immix; for each 64-kilobyte block used in our implementation of the
Nofl space, there are 4 kilobytes of metadata, whereas there would only
be 512 bytes of line marks in an Immix system.

\paragraph{Laziness}
Another difference is that in Nofl, sweeping is lazy: holes are
discovered during allocation by mutators, not eagerly after a trace. In
the best case, this is an efficient, cache-friendly source of
parallelism, naturally distributing work among mutators without
scheduling spikes. In the worst, Nofl will waste mutator time
re-scanning blocks full of survivor objects without significant holes.
The risk of run-time overhead is higher than with Immix, whose coarse
scan guarantees a minimal hole size, but on the other hand Nofl's
precision allows it to exhibit better fragmentation, as it does not have
Immix's pathological case where one small object can keep two 128-byte
lines alive.  We return to this point in section~\ref{sec:evaluation}.

\section{Implementation}

Before describing the collector implementation, we feel we must offer an
apologia: despite C's continued decline in favor of memory-safe
languages, we decided to write a new library in C. We can only plead
that we are targetting a low-dependency embedder whose purpose is to
enable more code in high-level, garbage-collected languages.

The embedder under consideration is the venerable Guile Scheme
implementation \cite{Guile}. Guile is itself mostly written in Guile
Scheme, but its runtime still has 130K lines of C source, with a C API
exposed to third-party users. Guile had its own bespoke garbage
collector in the distant past but has used the Boehm-Demers-Weiser (BDW)
collector \cite{BDW} since 2008, and as such there are many ways in
which conservative root-finding and even conservative heap-tracing are
embedded across its codebase; migration is possible over time, but given
the limited resources at our disposal, we needed to find a low-risk,
incremental path to better GC.  Immix offered interesting possibilities,
as it meant we could move towards evacuating collectors while allowing
targets of conservative roots to be marked in place, at least in the
near- and mid-term.

\subsection{Whippet: A Collector Collection}

Our work-in-progress effort is called
Whippet\footnote{\anon{\url{https://github.com/wingo/whippet}}}, and is a
no-dependency library designed to be embedded in a host run-time's
source tree.  Whippet defines a high-level abstract API which can be
implemented by a number of different collectors. Selection of the
concrete collector implementation is made at compile-time. Whippet
exposes enough detail to the embedder to allow open-coding of fast paths
for allocation, cooperative safepoints, and write barriers.

The concrete collectors included in Whippet are a simple serial
semi-space collector (\texttt{semi}), a parallel copying collector
(\texttt{pcc}), the Nofl-based mostly-marking collector (\texttt{mmc}),
and a shim that uses the BDW collector (\texttt{bdw}). We use
\texttt{semi} as an oracle to verify that \texttt{pcc} has the expected
performance characteristics, and similarly use \texttt{pcc} as an oracle
for \texttt{mmc}. \texttt{bdw} is included as a third-party comparison.

All collectors have thread-local allocation buffers (or free-lists in
the case of \texttt{bdw}).  \texttt{mmc} and \texttt{pcc} collect in
parallel and share an implementation of parallel tracing; newly shaded
grey objects go onto worker-local FIFO worklists, which overflow to
per-worker Chase-Lev deques \cite{chase2005dynamic, le2013correct},
allowing for fine-grained work-stealing of individual grey objects.
\texttt{bdw} has its own implementation of parallelism; \texttt{semi} is
serial.  All collectors but \texttt{semi} support multiple mutator
threads, and all are stop-the-world collectors, performing all trace
work during GC pauses.  \texttt{semi} and \texttt{pcc} align their
allocations on 8-byte boundaries, whereas \texttt{mmc} and \texttt{bdw}
use 16-byte alignment.

Whippet also includes a number of facilities needed for practical
garbage collector implementations:  embedder-provided root set
representations, trace functions, and forwarding mechanisms; three heap
sizing policies (fixed, grow-only heap size multipliers, and MemBalancer
\cite{kirisame2022optimal}); resurrecting finalizers; ephemerons; online
statistics; user-space tracepoints \cite{desnoyers2006lttng,
  ImmixTracing}; and so on.

\subsection{Nofl-Based Mostly-Marking Collector}

This paper's main focus is on the Nofl-based \texttt{mmc}. The
mostly-marking collector consists of a single Nofl space, augmented by a
mark-sweep large object space for allocations greater than 8192
bytes. \texttt{mmc} can be built in a number of configurations:
parallel or not, generational or not, and with varying levels of
precision: fully precise, with conservative roots but precise heap
tracing, or a fully-conservative configuration. Conservative edges
prevent evacuation of their referents, as we cannot be sure that the
source of the edge is actually an object reference
\cite{boehm1988garbage}; for roots-only conservative edges, this
prevents evacuation of the small subset of objects referenced from
roots. In this paper we focus on the parallel, fully precise,
non-generational configuration of \texttt{mmc}, as it it highlights the
characteristics of the Nofl space.

Memory for the Nofl space is reserved in units of 2-megabyte-aligned
slabs, divided into 64-kilobyte blocks. Nofl's minimum alignment
(granule size) on a 64-bit system is 16 bytes. There is a full byte of
metadata per granule. When an object is allocated, a 1 is written into
the metadata byte array, indicating the start of a recently-allocated
object, and a 32 is written into the byte corresponding to the last
granule of the object (or 33 for one-granule objects). There is a
current mark value, either 2, 3, or 4, and which rotates at each
collection cycle; marking replaces the low 3 bits of the object's
initial metadata byte with the current mark. Having a fixed mark for
young objects simplifies the allocation inline path, while also enabling
sticky-mark-bit generational collection \cite{demers1989combining}.

\begin{figure}
  \centering
  \includegraphics[width=\linewidth]{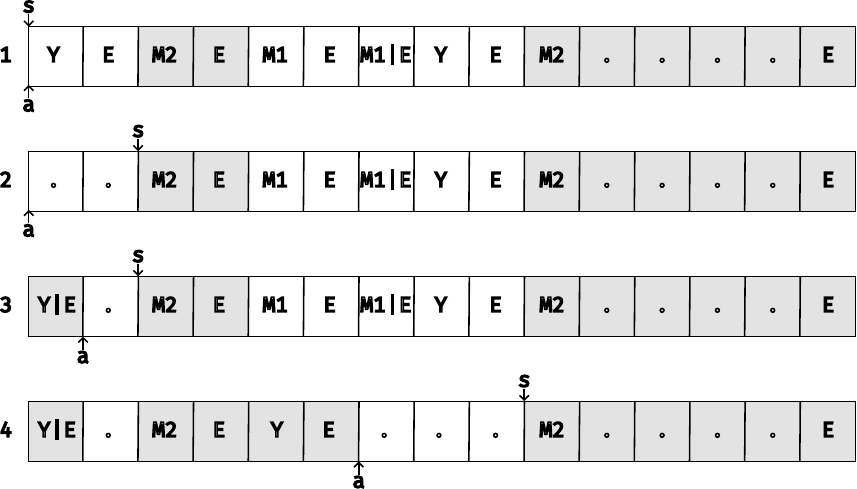}
  \caption{Lazy sweeping over the mark array of a Nofl block:  (1)
    Initially, the the sweep pointer \texttt{s} and the allocation
    pointer \texttt{a} are at the beginning of the block's mark table.
    (2) The allocator begins looking for a hole by scanning for a byte
    with the current mark (M2), or end-of-block.  It finds the
    end-of-hole at offset 2, so it advances \texttt{s} to 2 and clears
    the mark table between \texttt{a} and \texttt{s}.  (3) We allocate a
    single-word object, so we write the young and end-of-object markers
    at \texttt{a}, then advance \texttt{a} by one.  (4) We go to
    allocate a two-word object; the hole is too small, as
    \texttt{s}$-$\texttt{a} is 1.  Sweeping advances \texttt{s} over
    live objects marked with M2 by scanning for end-of-block, repeats as
    long as \texttt{s} points to a live object, sets \texttt{a} to
    \texttt{s} to start the hole, scans forward for end-of-hole as
    in (2) and allocates the object as in (3).}
  \label{fig:mark-table}
  \Description{Four illustrations of the sweep and allocation pointers
    for a block containing survivors marked with M2, old objects that became
    unreachable marked with M1, as well as newly allocated objects
    marked with Y that also became unreachable.}
\end{figure}

When a mutator obtains a block, it begins by sweeping it from the
beginning, looking for a hole. Sweeping scans for the first byte whose
low bits are the current mark, loading 8 metadata bytes at a time using
SIMD-within-a-register search techniques; any granules before the first
mark are in an allocatable hole. The granules corresponding to the hole
as well as the hole's metadata bytes are eagerly cleared when the hole
is discovered. When the hole is too small for allocation, the
fragmentation is added to a counter associated with the block, and
sweeping then skips over live objects by searching for a metadata byte
with the end bit (32) set, repeating the process if the next byte has
the live mark.  Figure~\ref{fig:mark-table} shows an example of this
process, with the block size reduced for clarity.

When attempting to allocate into a block that has been
fully swept and the last hole is finished, a mutator thread fetches a
fresh block from a global worklist and tries again.

As described above, there are 4 unused bits in the metadata byte, which
opens up interesting possibilities. In \texttt{mmc}, we have used the
byte to store bits for precise field-logging write barriers, per-object
generations, a bit indicating untagged allocations, and pinned
objects. The metadata byte array for blocks in the slab is accounted for
in the overall heap usage for \texttt{mmc}, and is stored in the
beginning of the slab, and so can be located with simple address
arithmetic.

The original \texttt{mmc} collectors that we built only had lazy
sweeping, but we found it advantageous to re-introduce eager sweeping
over empty blocks: as each object is marked, the collector ensures that
a mark is set on the block containing the object. This allows us to
eagerly determine which blocks are entirely free after collection. These
blocks can then be returned to the operating system, allocated into
without sweeping the mark byte array, or reserved for evacuation. Adding
eager block marks actually decreased total time for all benchmarks,
despite the additional highly-shared memory accesses. This is largely
because mutators did not have to themselves eagerly sweep blocks until
they found empties, when they needed to reallocate storage from the Nofl
space to the large object space.

\section{Evaluation}
\label{sec:evaluation}

\begin{table*}
  \caption{Microbenchmarks}
  \label{tab:microbenchmarks}
  \begin{tabular}{cclcc}
    \toprule
    Benchmark name&Parameters&Origin&Minimum heap (MB)&Total allocation (8/16 byte alignment)\\
    \midrule
    \texttt{nboyer} & 4 & Boyer/Clinger/R7RS & 70 & 758 / 786\\
    \texttt{nboyer} & 5 & Boyer/Clinger/R7RS & 208 & 2252 / 2307\\
    \texttt{peval} & 12, 1 & Feeley/R7RS & 37 & 2290 / 2409\\
    \texttt{gcbench} & 0 & Ellis/Kovac/Boehm & 20 & 597 / 715\\
    \texttt{splay} & 10 & Octane & 47 & 1525 / 1609 \\
    \texttt{earley} & 14 & Feeley/R7RS & 250 & 2290 / 2983 \\
  \bottomrule
\end{tabular}
\end{table*}

\subsection{Whiffle: A Scheme to Test Whippet}
While the ultimate goal of this effort is to replace the memory manager
in Guile, Guile itself is not currently a good workbench for garbage
collector evaluation. In particular, because Guile relies on
conservative root-finding, we are unable to use Guile to compare a
Nofl-based collector to the more well-understood performance profile of
a copying collector.

Instead we decided to implement a new Scheme-to-C compiler, with the
express goal of testing and evaluating Whippet. This ``Whiffle'' Scheme
implementation\footnote{\anon{\url{https://github.com/wingo/whiffle}}} re-uses
the reader and syntax expander from Guile, applies Guile's
\texttt{peval} pass implementing Waddell and Dybvig's inlining algorithm
\cite{waddell1997fast}, and then emits C directly in the style of a
baseline compiler. Each thread has its own managed stack which it uses
to store named and temporary values. These values are allocated with a
stack discipline, and recording a safepoint requires only storing the
managed stack pointer. Scheme implements iteration by tail calls, and
Whiffle actually compiles these as such, relying on the C compiler to
compile tail calls as jumps. The generated code explicitly checks for
safepoints at every function entry.

Compared to a production Scheme implementation, Whiffle allocates many
more closures due to a lack of contification and closure optimization
\cite{fluet2001contification, keep2012optimizing}, and otherwise does
not do any unboxing. All Whiffle values are tagged:  characters,
booleans, null, and small integers are immediate, and all others are
heap objects with tags in the low bits of their first word. Careful
selection of heap tags allows pairs to be represented in only two words.
This representation strategy is similar to what is used in Guile, making
Whiffle a reasonable proxy for how Whippet will perform when integrated
into Guile, while also allowing comparison with precise collectors.

\subsection{Methodology}
Our initial evaluation is limited to the microbenchmarks listed in
Table~\ref{tab:microbenchmarks}.  \texttt{nboyer}, \texttt{peval}, and
\texttt{earley} are standard Scheme microbenchmarks, ported from the
R7RS version of the Larceny benchmark
suite\footnote{\url{http://www.larcenists.org/benchmarksGenuineR6Linux.html}}.
\texttt{splay} is a port from the Octane 2.0 JavaScript
benchmark\footnote{\url{https://github.com/chromium/octane}}.
\texttt{gcbench} is a port the benchmark distributed at
\url{https://www.hboehm.info/gc/gc_bench.html}, adapted to run in
multiple threads, optionally inserting fragmentation between
allocations; passing 0 as a parameter, as we do in these tests, uses the
traditional behavior with no fragmentation.

For all of these benchmarks, we test at a range of heap sizes, expressed
as multiples of the minimum size at which we could get the benchmark to
run, minus the known overhead of the collector on which the minimum run
succeeded. All of these benchmarks are serial; to simulate load from
multiple mutator threads, we run multiple instances of the benchmarks in
parallel threads, and scale the heap accordingly. For example, the
\texttt{nboyer-5} benchmark with a 2.5$\times$ heap and eight mutators
will be run with a heap size of 4160 megabytes.  Incorporating workloads
that involve inter-thread communication and more realistic shared heap
behavior is left to future work.

We measured the performance of single-threaded \texttt{semi} against the
more elaborate parallel copying collector (\texttt{pcc}) configured in a
single-mutator, single-worker configuration.  The results for
\texttt{pcc} and \texttt{semi} are similar enough to omit further
discussion in this paper. In the rest of this section, we focus on
\texttt{pcc}, which implements copying collection in a way that scales
to multiple mutator and collector threads.

Out of the many configurations available to the mostly-marking
collector, we focus on the parallel, precise, non-generational
configuration. We abbreviate this configuration as \texttt{mmc} in the
results below.

We run our tests on an AMD Ryzen Threadripper PRO 5955WX system (16
cores, 32 logical CPU threads, 32 MB last-level cache, 4/4.5 GHz) with
128 GB of DDR4-3200MHz RAM, running Guix GNU/Linux at commit
90ee330bafc5a954 (February 2025). No use of \texttt{isolcpus}, thread
pinning, boost disabling, or other specific performance tuning has been
made on the test system. Each test was run 10 times; in the graphs
below, we plot all points and paint a median line for each scenario.

We apply the lower-bound overhead (LBO) methodology
\cite{cai2022distilling} to measure the cost of garbage collection. For
a given benchmark, mutator count, and GC worker count, we identify the
result with the lowest time attributable to the mutator, and take it to
be a conservative approximation of the time that the benchmark would
take without GC overhead (the minimum distilled cost). By dividing the
total time for each data point by the minimum distilled cost, we obtain
a lower bound of the overhead of garbage collection in that
configuration.

\subsection{Results}
\begin{figure*}
  \centering
  \includegraphics[width=\linewidth]{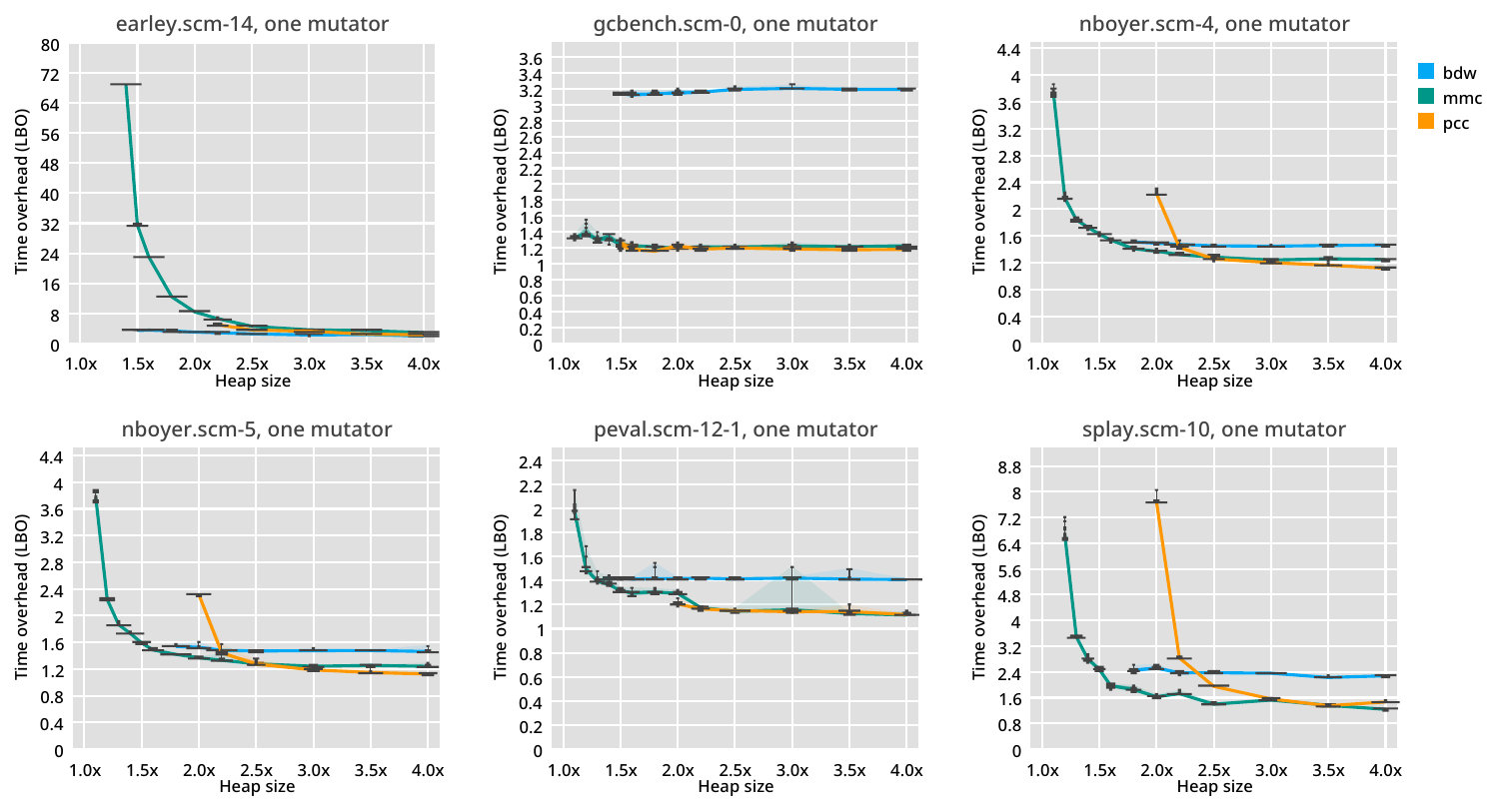}
  \caption{Wall-clock time overheads imposed by the different collectors
    for benchmarks run with a single mutator thread.  The vertical axes
    show lower-bound overheads (LBO): a ratio of total time divided by
    the minimum observed time to complete the benchmark, not counting GC
    pauses.}
  \label{fig:elapsed-lbo-1}
  \Description{Six graphs, one per benchmark, showing decreasing
    overhead as heap sizes grow.  In 5 of 6 configurations, \texttt{mmc}
    is lowest-overhead for tight heap sizes, and is sometimes bettered
    by \texttt{pcc} at generous heap sizes.  \texttt{earley} is an
    anomaly, in which \texttt{mmc} overhead is higher than
    \texttt{bdw}.}
\end{figure*}
\begin{figure*}
  \centering
  \includegraphics[width=\linewidth]{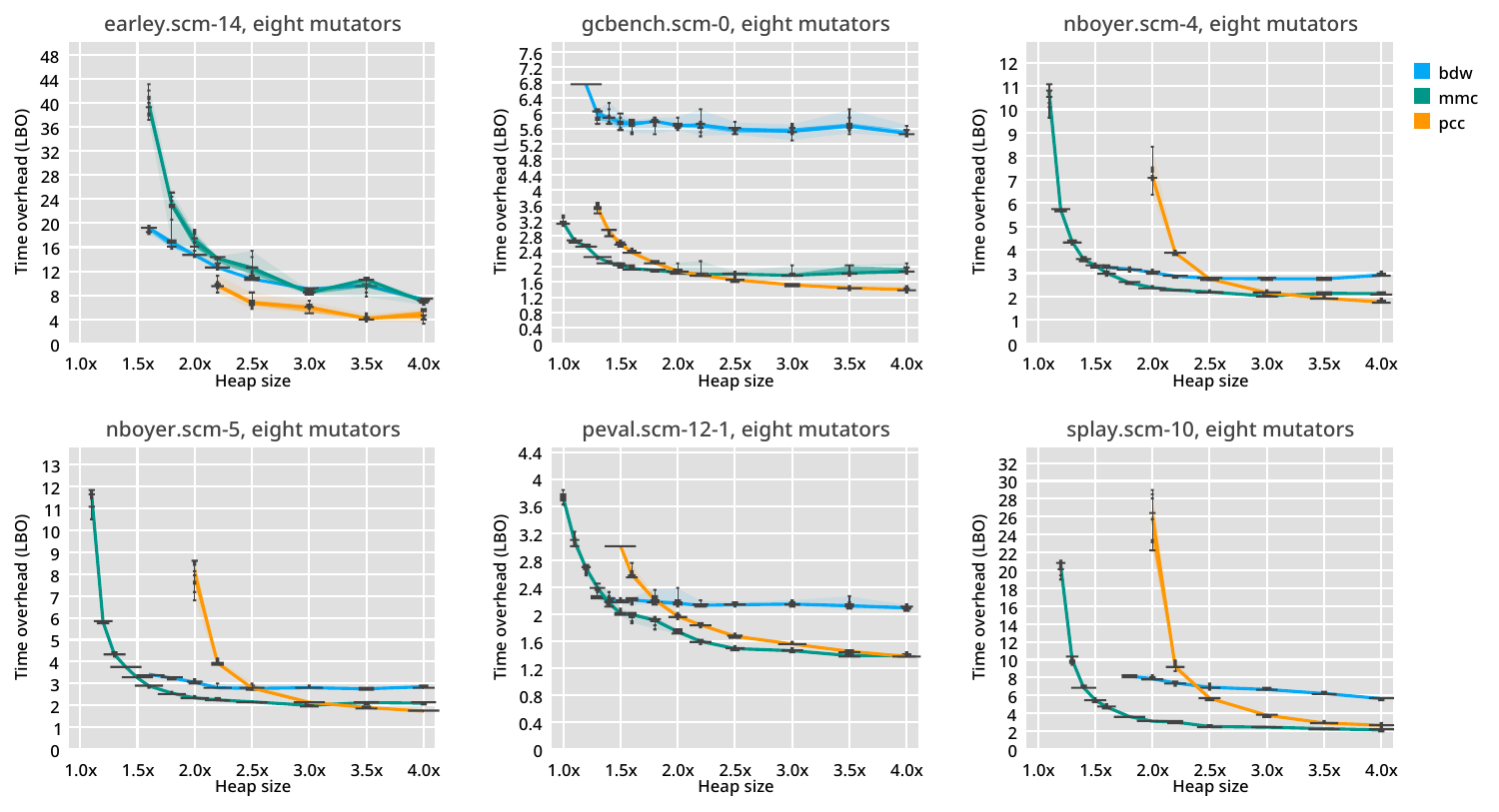}
  \caption{Wall-clock time overheads as in
    Figure~\ref{fig:elapsed-lbo-1}, but with eight mutator threads and
    heap sizes scaled up by eight.}
  \label{fig:elapsed-lbo-8}
  \Description{Six graphs in the same configuration as
    Figure~\ref{fig:elapsed-lbo-1}.}
\end{figure*}

\begin{figure*}
  \centering
  \includegraphics[width=\linewidth]{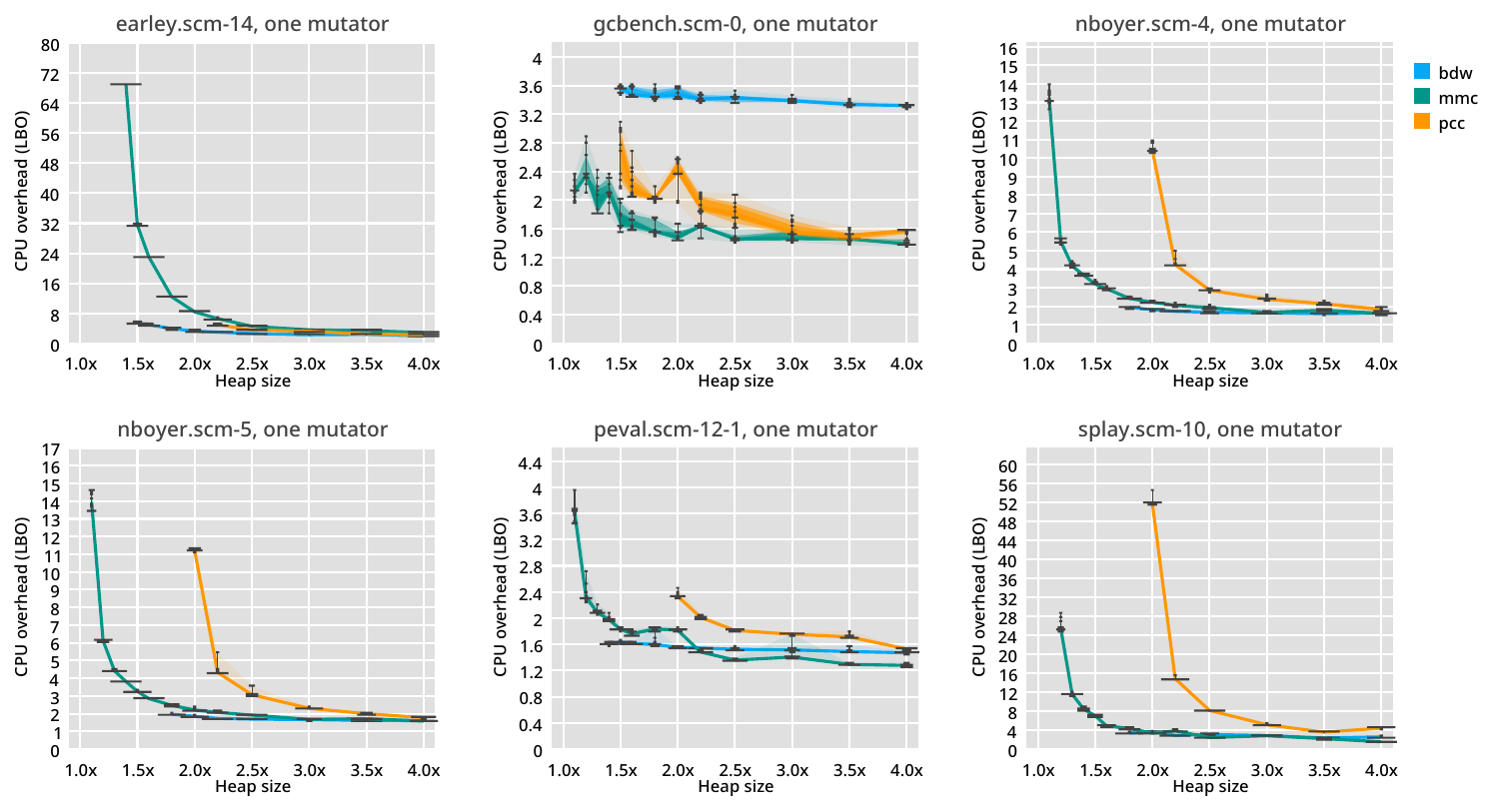}
  \caption{CPU time overheads, single mutator thread.  Compare to
    Figure~\ref{fig:elapsed-lbo-1} which measures wall-clock time
    instead.}
  \label{fig:cpu-lbo-1}
  \Description{Six graphs in the same configuration as
    Figure~\ref{fig:elapsed-lbo-1}.}
\end{figure*}

\begin{figure*}
  \centering
  \includegraphics[width=\linewidth]{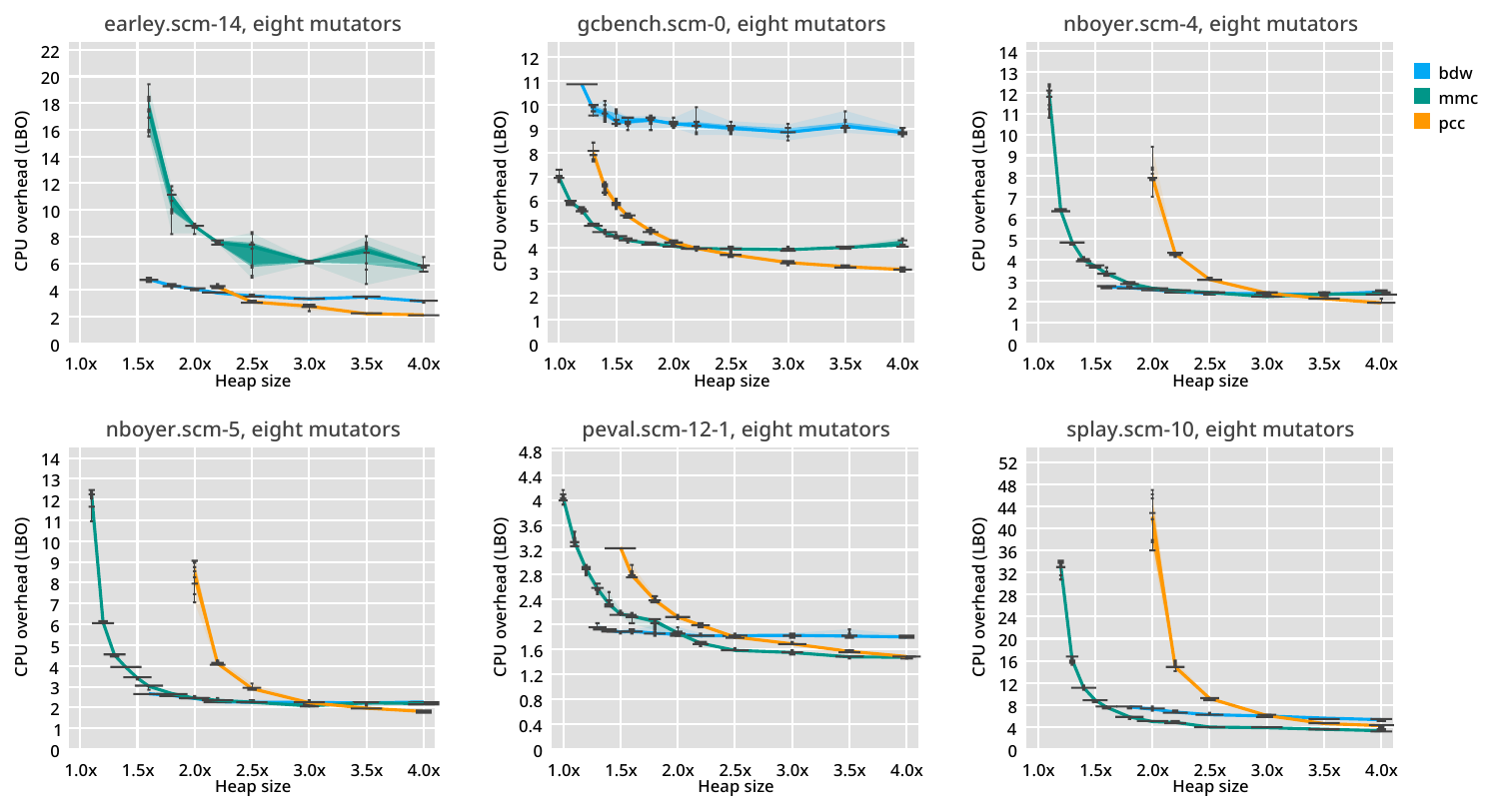}
  \caption{CPU time overheads, eight mutator threads.  Compare to
    Figure~\ref{fig:elapsed-lbo-8} which measures wall-clock time
    instead.}
  \label{fig:cpu-lbo-8}
  \Description{Six graphs in the same configuration as
    Figure~\ref{fig:elapsed-lbo-8}.}
\end{figure*}

\subsubsection{Single mutator, wall-clock time}

Figure~\ref{fig:elapsed-lbo-1} shows the lower-bound wall-clock time
overhead for \texttt{mmc}, \texttt{bdw}, and \texttt{pcc} for the six
microbenchmark configurations above, running a single mutator thread and
8 collector threads.  For this analysis and the ones that follow, we
present our conclusions as a list of enumerated observations.

\begin{enumerate}
\item \texttt{pcc} performs as expected for a copying collector.  It
  runs in less than 2.0$\times$ on \texttt{gcbench} because of allocations in
  the large-object space, which does not need a copy reserve.
\item For all tests, the Nofl-based \texttt{mmc} collector can run in
  tight heaps where \texttt{bdw} and \texttt{pcc} fail to run.  It does
  so at high overhead, however, and for \texttt{earley} on tight heaps,
  \texttt{mmc} takes longer than the 90-second timeout of our tests.
\item \texttt{mmc} performs poorly on \texttt{earley} unless the heap
  size is very generous.  We discuss this in more detail below.
\item \texttt{bdw} overhead is only weakly sensitive to heap size.  We
  find this to be unexpected.
\end{enumerate}

\subsubsection{Eight mutators, wall-clock time}

Figure~\ref{fig:elapsed-lbo-8} is as above, but for 8 mutator threads
instead of 1, with total heap sizes multiplied by 8 as well.

\begin{enumerate}
\item Multi-mutator overheads are generally higher than single-mutator
  overheads.  We attribute this to the costs of tracing a scaled-up
  heap: while allocation parallelizes well, collection has to trace a
  larger heap with the same number of workers.
\item The exception is again \texttt{earley}, on which \texttt{mcc}
  still performs poorly, but is less of an outlier relative to the
  single-threaded case.
\end{enumerate}

\subsubsection{CPU time}

Figures~\ref{fig:cpu-lbo-1} and~\ref{fig:cpu-lbo-8} show overhead in
terms of measured CPU time instead of wall-clock time.

\begin{enumerate}
\item Though their wall-clock overheads are markedly different, the CPU
  overhead of \texttt{bdw} and \texttt{mmc} are essentially identical
  for these benchmarks.  Looking at the data, we see that this is
  because \texttt{bdw} does not parallelize very well during tracing;
  for example, the ratio of CPU time to wall-clock time during GC pauses
  in \texttt{nboyer-4} is 5.4$\times$, whereas for \texttt{mmc} it is
  7.6$\times$ and perfect speedup would be 8.0$\times$.
\item It is a well-known result, but it is still striking to see that
  giving a process 10\% more heap can cut its CPU overhead in half, at
  the tight end of the curve.
\end{enumerate}

\subsubsection{Discussion and future directions}

We are unable to make proper conclusions based on microbenchmarks alone;
a full evaluation will require testing against a production language
run-time with real work-loads.  However, the limited results above do
provoke some thoughts.

\paragraph{The case of \texttt{earley}}

Relative to the other microbenchmarks, \texttt{earley} exhibits
anomalous behavior for \texttt{mmc}.  Looking closer into the raw data,
we notice that CPU utilization is very low; for example, at 2.5$\times$
heap with one mutator, \texttt{mmc} spends 1.4 wall-clock seconds in the
mutator and 2.5 wall-clock seconds stopped in GC pauses, but CPU time
during GC pauses is also only 2.5s, for a CPU utilization of
1.0$\times$.  This indicates that the pause is not parallelizing.

Looking at the other collectors, we see pause-time CPU utilization in
\texttt{bdw} of 1.1$\times$, which is also low.  \texttt{pcc} exhibits
the same behavior as \texttt{mmc} with a CPU utilization of 1.0$\times$.
Because of the low CPU utilization during pauses for all collectors, we
conclude that the graph is not parallel.  However, it is also clear that
there is also a performance pathology, given that at a 1.5$\times$ heap
multiplier, \texttt{mmc}'s CPU LBO is six times higher than that of
\texttt{bdw}.  Further investigation is needed.

More generally, we observe that the LBO methodology appears to
effectively summarize overall GC costs, but is insufficient to identify
the structure of the overhead.  We had to look at the absolute
measurements that comprised CPU-time LBO to make a conclusion about
graph parallelism, and to actually identify the pathology will require
more lower-level data.

\paragraph{Sediment}
While \texttt{mmc} can run in very tight heaps, its overhead is high in
these configurations.  Some of this appears to be unavoidable in a
whole-heap collector; there is no getting around the need to trace the
whole graph.  As the number of garbage collections is proportional to
the inverse of the available space, each halving of the available space
will lead to twice the number of collections, and thus twice the
overhead.

However, we speculate that some of the lazy-sweep work may be avoided.
Some blocks will be populated by long-lived objects and will have little
fragmentation; by avoiding allocating into these blocks, we may avoid
the need to sweep them also.

In our implementation of the Nofl space, we use the conventional
strategy of rotating the current mark at each collection, to avoid
having to clear marks on live objects.  After each collection, a block
is composed of survivors and holes.  For the side table of mark bytes,
these holes consist of spans of dead objects, which start with a mark
byte that is not current, and fragmentation, whose mark bytes are zero.
In one sense, the contents of a hole do not matter, as they do not
contain reachable objects.  However when the mark value rotates around
again, if hole mark bytes are not cleared, dead objects would be seen by
an allocator to have the live mark, preventing their memory from being
re-used.  Therefore, sweeping \emph{should} clear holes.  For a Nofl
space in a collector with conservative roots, sweeping \emph{must} clear
holes, to prevent dead objects from being resurrected.

In summary, a Nofl block needs sweeping if it contains dead objects.
Conversely, a full block that contain only survivors is not a good
allocation target; as future work, we should consider collecting more
per-block statistics during tracing and sweeping that would allow us to
identify these \emph{sedimentary} blocks and avoid allocating into them.

By way of comparison, we note that Immix implicitly has the sedimentary
optimization: it avoids allocating into blocks with no free lines.
However in a configuration with conservative roots, we do not see how it
can avoid a similar fine-grained sweep over all objects, unless it too
keeps statistics to identify blocks whose objects all survived.

Another direction would be to sweep concurrently instead of lazily;
sweeping a block could collect its holes into a freelist, threaded
through the first word of each hole.  There are enough bits to also
include hole size.

\paragraph{Fragmentation}
How can we expect that sweeping a block will actually find good-sized
holes?  To an extent we have avoided the question up to now, instead
relying on Immix's evacuation heuristics to detect blocks with high
fragmentation and to evacuate them.  However, we need to measure.
Future work should record the distribution of live object and hole
spans, as well as the distribution of holes that any allocation had to
sweep.  Immix had to add an \emph{overflow allocator} for medium-sized
objects to optimize this case; perhaps Nofl will run into something
similar, as it is applied to production workloads.

\paragraph{The allocation sequence}
Allocating an object in the Nofl space currently also writes into the
mark table.  Having object extent recorded in the mark table is useful,
as it makes for a cheap sweep.  However, most objects die young; in
precise configurations, which do not need to query whether an address
references an object or not, Nofl could instead record object extent
when an object is first traced.  This would avoid a write during the
allocation sequence, and avoid a write entirely for an object that did
not survive the first collection.  It would be valuable to prototype
this behavior, to measure the overhead of the side-table writes during
allocation.

\paragraph{Production collection}
Another way to approach the benchmarking problem is to target a
different language entirely.  While we have designed Whippet for Guile's
use case, it is completely independent; embedding Whippet into another
language run-time, for example OCaml, would allow for evaluation against
that language's benchmark suite.  It would also allow for comparison of
Whippet's collectors against the new target's own collectors, which
would accelerate progress in both.

\section{Related Work}

Nofl's target use case is Guile \cite{Guile}, which currently uses the
Boehm-Demers-Weiser collector
\cite{boehm1988garbage,boehm1993space,BDW}.  BDW is a mature collector
that offers a useful point of comparison for performance.  We show that
the Nofl design point would seem to be a net performance improvement,
relative to BDW; Nofl parallelizes better, both during allocation and
collection, it can run in tighter heaps for a given overhead, and its
optimistic evacuation can solve the thorny fragmentation problem that
plagues all mark-sweep collectors, BDW included.

The Nofl design is an evolution of Immix \cite{Immix08}, exploring what
Immix would be like if its lines were shrunk to a minimum.  We are
unable, however, to compare Nofl to Immix directly.  This is because
collector construction is not a simple continuous function of line size:
once lines become precise marks, one wants to build the collector in
qualitatively different ways, for example encoding extent precisely,
using bits of mark bytes for other purposes, and of course not needing
separate mark tables and line tables, not to mention object map,
pinning, and field-logging bits.  Still, it would be illuminating to
build separate Immix and Nofl collectors into the same language
run-time.  In general, like-to-like comparison is fraught: if we built
Immix into Whippet, there is the risk that our implementation would
somehow not be representative; if we built a Nofl space into an
Immix-using system like OpenJDK with MMTk, the design space is different
in that OpenJDK reserves space for GC state within object
representations, whereas Guile and Whiffle do not.  Still, any efforts
towards direct comparison of Immix with Nofl would be of value.

Nofl is specifically inspired by Conservative Immix and Sticky Immix
\cite{shahriyar2014fast, demers1989combining}.  We consider Conservative
Immix to offer a new hope for language run-times that do not currently
record precise stack roots, as it pins referents of ambiguous
(conservative) roots, but allows evacuation of all other objects.  It
solves the ``does this address reference an object'' problem by
maintaining an object map, containing a bit for every potential object
start address.  The object map bit is set for every object that survived
the previous collection or which has been allocated since, is cleared
after conservative roots are scanned, and rebuilt at each trace.  Nofl
uses the side table of mark bytes instead as an object map.  Further
evaluation of conservative Nofl configurations are left to future work.

The idea of sequential allocation into holes appears to have originated
with Dimpsey et al's parallel mark-sweep/mark-compact collector for
IBM's server JVM \cite{Dimpsey2000}.  This collector finds worst-fit
spans from a global free-list, splits them into chunks approximately
1024 bytes in size, and uses those split chunks for thread-local
bump-pointer allocation of small objects.  It also represents mark bits
in a side table, and sweeps over that table instead of the object
memory.  Immix builds on this strategy, and a number of collectors have
followed since, for example the concurrent mark-region collector in
Schism \cite{pizlo2010schism} or the Dartino collector \cite{Dartino}.
Unlike Immix, we do not use the term \emph{mark-region} to describe the
Nofl algorithm, as we feel that the regions (lines) in question are so
small as to not be worth emphasis.  Instead we see Nofl as closer to
mark-sweep collectors, but with bump-pointer allocation, and built to
also allow for Immix-style optimistic evacuation.

The Whippet collector library that we built is similar to MMTk
\cite{blackburn2004oil, mmtk}.  It makes different design trade-offs due
to its target users, specifically Guile, but also other production
language run-times implemented in C.  Whippet takes direct inspiration
from MMTk's API, its use of data types to prevent confusion between
different ways addresses are used, and its internal organization.  It
was specifically inspired by the initial Rust Immix work
\cite{RustImmix}, which uses a Chase-Lev deque for work-stealing
\cite{chase2005dynamic}; Nofl uses the version of Chase-Lev deques that
is adapted for the C11 memory model \cite{le2013correct}.  Since the
Rust Immix publication, MMTk itself has since switched its own
implementation to Rust, which also reworking its unit of parallelism
into work packets \cite{ossia2002parallel} instead of more ad-hoc,
fine-grained mechanisms such as classic work stealing.  We have not yet
measured single-object work-stealing to be a bottleneck in our testing
with up to 8 trace workers, but this may become an issue in the future.

We have implemented user-space tracepoints in Whippet, as Huang et al
did in MMTk \cite{ImmixTracing}, though we used LTTng instead of EBPF.
We confirm that it has been useful in identifying the causes of
performance problems, for example helping resolve an issue in detecting
termination of work-stealing.  We suspect that it will be useful when we
go to identify the pathology preventing good performance on the
\texttt{earley} benchmark.

Another similar recent work is Patton's Immix-based parallel collector
for SBCL \cite{patton2023parallel}, which also uses conservative
root-finding.  Like MMTk, Patton's collector uses the grey-packet design
of Ossia et al.  Patton's collector also uses an interesting trick to
lazily initialize the mark table at trace time: lines allocated into
during the previous cycle are marked as \emph{fresh}, and a conservative
reference into a fresh line is resolved by finding the start of the line
and then iterating forward.  It would be fruitful to compare this
approach to Nofl's eager object map initialization.

\section{Conclusions}
Immix's design suggests that collectors should reclaim memory with fine
granularity, but no finer than a line.  But what if we did go as fine as
possible?  This paper finds that precise reclamation is perfectly
workable, and merits further investigation.  Our Nofl design
incorporates granule-level reclamation.  We build an implementation of
Nofl in a new embeddable garbage collection library, Whippet, and a new
Whiffle Scheme-to-C compiler as a benchmarking workbench.  Comparing
against our parallel copying collector and the third-party
Boehm-Demers-Weiser collector, we find that in terms of wall-clock time,
the Nofl-based mostly-marking collector outperforms the others for
anything other than a very generous heap, with the exception of one
benchmark (\texttt{earley}).  We discuss the possible reasons for this
behavior and outline future avenues for research.

\begin{acks}
The authors would like to thank CF Bolz-Tereick for helpful discussions
and feedback.

The authors would also like to thank the anonymous referees for their
valuable comments and helpful suggestions.  This work was supported by
\grantsponsor{nlnet-whippet-2024}{NLnet
  Foundation}{https://nlnet.nl/project/Whippet/} as part of EU
Next-Generation Internet grant
\grantnum[https://cordis.europa.eu/project/id/101092990]{nlnet-ngi0-2022}{101092990}.
\end{acks}

\bibliographystyle{ACM-Reference-Format}
\bibliography{wingo-bibliography}

\end{document}